\documentclass[aps,prl,twocolumn,floatfix,superscriptaddress,showpacs]{revtex4}
\usepackage{graphicx}
\usepackage{mathrsfs} %\mathscr 
\usepackage[mathscr]{euscript}
\usepackage{amsmath,amssymb}
\usepackage{bm}
\usepackage{color}

\def\ia{{\accent 19 \char 16}}
\def\i{{\bf i}}
\def\Tr{{\rm Tr}}
\def\ket#1{ | #1 \rangle }

\def\braket#1#2{ \langle #1 | #2 \rangle }

\def\vec#1{ {\bm{#1}} }

\def\expe_av#1{ \left \langle #1 \right \rangle_{\rm av}} 
\newcommand{\1}{\mbox{1}\hspace{-0.25em}\mbox{l}} %
\definecolor{purple}{rgb}{0.62745098,0.125490196,0.941176471}
\definecolor{darkgreen}{rgb}{0,0.6,0}
\begin{document}

%\preprint{}

\title{Symmetry-protected topological phases and transition in a frustrated spin-$\frac{1}{2}$ XXZ chain}


\author{Hiroshi Ueda}
\affiliation{Condensed Matter Theory Laboratory, RIKEN, Wako, Saitama 351-0198, Japan}
\author{Shigeki Onoda}
\affiliation{Condensed Matter Theory Laboratory, RIKEN, Wako, Saitama 351-0198, Japan}
\affiliation{RIKEN Center for Emergent Matter Science (CEMS), Wako, Saitama 351-0198, Japan}

\date{\today}

\begin{abstract}
Frustrated spin-$1/2$ XXZ zigzag chains relevant to Rb$_2$Cu$_2$Mo$_3$O$_{12}$ are revisited in the light of symmetry-protected topological (SPT) phases. Using a density-matrix renormalization group method for infinite systems, we identify projective representations for four distinct time-reversal invariant SPT phases; two parity-symmetric dimer phases near the Heisenberg and XX limits and two parity-broken vector-chiral (VC) dimer phases in between. A small bond alternation in the nearest-neighbor ferromagnetic exchange coupling induces a direct SPT transition between the two distinct VC dimer phases. It is also found numerically that two Berezinskii-Kosterlitz-Thouless transitions from the gapless to the two distinct gapped VC phases meet each other at a Gaussian criticality of the same Tomonaga-Luttinger parameter value as in the SU(2)-symmetric case.
\end{abstract}

%\keywords{}

\maketitle

Topological orders and the quantum entanglement provide novel notions for classifying gapped quantum states beyond the conventional Landau theory~\cite{Chen10}. These notions are indispensable for distinguishing between gapped ground states of the same symmetry group that are not adiabatically connected.
The entanglement remains short-range if the gapped ground state can be described as a direct (and thus unentangled) product of wavefunctions of finite-size blocks, and is long-range otherwise~\cite{Chen10}. Long-range entangled states may show nontrivial long-range topological orders either without any spontaneous symmetry breaking, as in $Z_2$ quantum spin liquids~\cite{Wen02}, or with a symmetry breaking, as in topological superconductors~\cite{Hasan10}. Short-range entangled (SRE) states can be transformed into each other without closing the energy gap. However, this transformation may necessarily break a certain symmetry. Then, this symmetry protects a topological distinction between the two SRE states. Such phases are referred to as symmetry-protected topological (SPT) phases. Well-known examples include the Haldane phase~\cite{Haldane83,Affleck87,Pollmann10,Pollmann12} of spin-1 chains having the time-reversal, dihedral, inversion symmetries and time-reversal invariant topological insulators~\cite{Hasan10}. The topological structure of an SPT phase with a symmetry group $G$ is characterized by an algebra of the projective representation of $G$ for the SRE ground state, and can thus be classified according to the group cohomology~\cite{Liu11,Chen11,Gu13,Schuch11}. Some one-dimensional (1D) interacting cases including the Haldane spin chain~\cite{Pollmann10,Pollmann12} and spin-$1/2$ ladders~\cite{Liu12} have been demonstrated numerically. 

However, a topological transition between distinct nontrivial SPT phases has not been reported yet in spin systems. This motivates us to study a simple yet more nontrivial case of a frustrated spin-$1/2$ chain~\cite{Chubukov91,Lecheminant05} 
including nearest-neighbor (NN) ferromagnetic ($J_1<0$), second-neighbor antiferromagnetic ($J_2>0$) exchange couplings, the relative amplitude of the NN bond alternation ($\delta$), and the XXZ-type easy-plane exchange anisotropy ($\Delta$);
\begin{eqnarray}
\hat{ \mathscr{H} }
 &=& J_1 \sum_{i} ( 1 - (-1)^i \delta )
\left[\hat{S}^x_{i} \hat{S}^x_{i+1}+\hat{S}^y_{i} \hat{S}^y_{i+1} + \Delta \hat{S}^z_{i} \hat{S}^z_{i+1}\right] \nonumber \\
&&+ J_2 \sum_{i} \left[\hat{S}^x_{i} \hat{S}^x_{i+2} + \hat{S}^y_{i} \hat{S}^y_{i+2} + \Delta \hat{S}^z_{i} \hat{S}^z_{i+2}\right],
\label{eq:H}
\end{eqnarray}
with a spin-$1/2$ operator $\hat{\bm{S}}_i$ at a site $i$.
Equation (\ref{eq:H}) with $\delta=0$ provides a minimal model for understanding the emergence of a long-range order (LRO) of the vector spin chirality, $\langle \hat{\kappa}^z\rangle = \frac{1}{N} \sum_i\langle (\hat{\vec{S}}_i \times \hat{ \vec{S} }_{i+1})^z \rangle \ne 0$ with $N$ being the number of spins~\cite{Nersesyan98,Hikihara01,FSO10,FSOF12}, and the associated ferroelectric polarization in various quasi-1D spin-$1/2$ cuprate Mott insulators~\cite{FSO10,Masuda05,Park07,Enderle05,Naito07,Yasui11,Wolter012}. A vital role of nonzero $\delta$~\cite{UO14} has been proposed for a gapped vector-chiral (VC) dimer state without a quasi-LRO of a spin spiral, in accordance with experiments on Rb$_2$Cu$_2$Mo$_3$O$_{12}$ which has a weak crystallographic dimerization~\cite{Hase04,Yasui13}. This induces two pairs of time-reversal and translation invariant gapped phases with and without the inversion symmetry, each pair of which belong to the same symmetry group~\cite{UO14} but are expected to possess a distinct topology protected by symmetries.

In this Letter, using the infinite-size density matrix renormalization group (iDMRG)~\cite{iDMRG} method, we classify these four gapped phases of this $J_1$-$J_2$ frustrated spin-$1/2$ XXZ chain model in terms of SPT phases. We also analyze the criticality of an SPT transition between two VC dimer phases, which supports the conformal field theory (CFT)~\cite{CFT96} of the central charge $c=1$.

\begin{figure} [H!tb]
\begin{center}
\includegraphics[width=8.6cm]{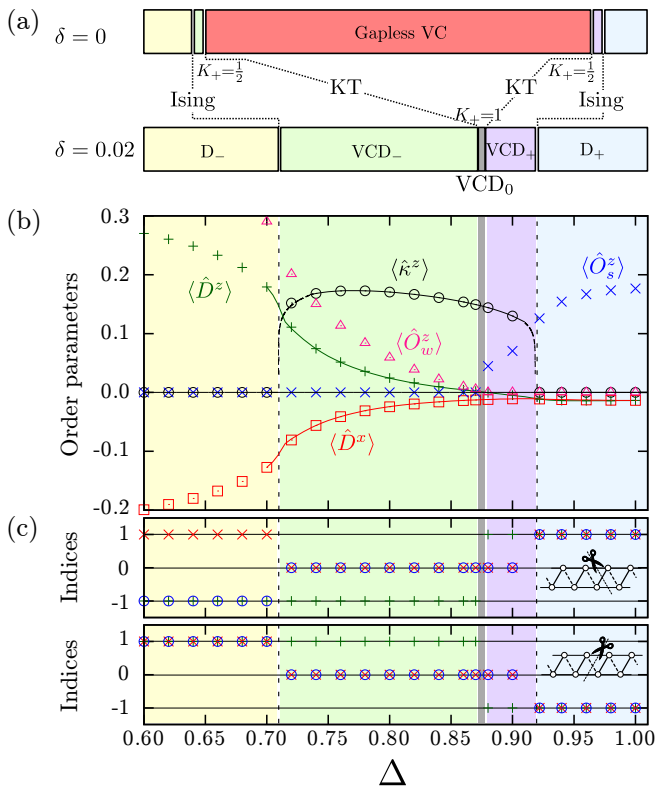}
\end{center}
\caption{ 
(Color online) 
Our iDMRG results for $J_1/J_2=-2.5$ obtained with 300 renormalized basis states ($m=300$). 
(a) Phase diagrams of the Hamiltonian given by Eq.~\eqref{eq:H} for $\delta = 0$ and $\delta = 0.02$.
(b) Order parameters for $\delta = 0.02$. Black, green, and red solid lines are extrapolations of the DMRG to the thermodynamic limit~\cite{UO14}.
(c) Variations in $Z_2$ indices [ $\beta(\Theta)$ and $\gamma(\Theta,R^{~}_{2z})$:\textcolor{darkgreen}{+}; $\beta(I)$:\textcolor{red}{$\times$}; $\omega(R^{~}_{2x} R^{~}_{2z})$:\textcolor{blue}{$\bigcirc$} ]. (Definitions are given in the text.) The upper/lower panel shows results obtained by dividing the system at a strong/weak $J_1$ bond. 
}
\label{fig:phasediagram}
\end{figure}

The ground-state phase diagram of Eq.~\eqref{eq:H} was revealed numerically in a wide range of parameters $\Delta$ and $J_1/J_2$ for $\delta=0$~\cite{FSO10,FSOF12} and $\delta\ne0$~\cite{UO14} and has also been reproduced by our present iDMRG calculations. In particular, the following distinct ground states appear with decreasing $\Delta$ from unity to zero for $-2.7\lesssim J_1/J_2\lesssim-1.5$~\cite{FSOF12}, as shown in Fig.~\ref{fig:phasediagram}~(a) for $\delta=0$ and $|\delta|=0.02$, with $J_1/J_2=-2.5$ being fixed.

i) Haldane dimer (D$_+$) state~\cite{FSOF12} ---
This is given by a Haldane state~\cite{Affleck87,Haldane83} of the NN spin pairs that are ferromagnetically coupled with the stronger relative amplitude $1+|\delta|$~\cite{FSOF12,UO14,Itoi01}. In this phase, two dimer order parameters $\langle \hat{D}^x\rangle=\langle \hat{D}^y\rangle$ and $\langle \hat{D}^z\rangle$ have the same sign while the vector spin chirality vanishes, i.e., $\langle\hat{\kappa}^z\rangle=0$, as shown in Fig.~\ref{fig:phasediagram}~(b) for $\delta=0.02$, where $\langle\hat{D}^\alpha\rangle=\frac{1}{N}\sum_i(-1)^{i-1}\langle\hat{S}^\alpha_{i}\hat{S}^\alpha_{i+1}\rangle$.

ii) Vector-chiral Haldane dimer (VCD$_+$) state ---
The state preserves the relation $\langle \hat{D}^x \rangle \langle \hat{D}^z \rangle >0$, while the parity symmetry is spontaneously broken by a LRO of the vector spin chirality; $\langle \hat{\kappa}^z \rangle \ne0$. 

iii) Vector-chiral dimer (VCD$_-$) state ---
This is similar to the VCD$_+$ state, except the sign of $\langle \hat{D}^z\rangle$ is reversed and thus $\langle \hat{D}^x\rangle\langle \hat{D}^z\rangle<0$.

iv) Gapless vector-chiral states---
The $z$-component dimer order parameter vanishes, $\langle \hat{D}^z\rangle=0$, while the LRO of vector spin chirality survive, i.e. $\langle \hat{\kappa}^z \rangle \ne0$. The other components $\langle\hat{D}^x\rangle=\langle\hat{D}^y\rangle$ are zero for $\delta=0$ (gapless VC phase)~\cite{Nersesyan98,Hikihara01,FSO10}, but are finite for $\delta\ne0$ (critical VCD$_0$ state)~\cite{UO14}. For $\delta\ne0$, the condition of $\langle\hat{D}^z\rangle=0$ for the VCD$_0$ state is satisfied only at a single direct transition point between VCD$_\pm$ phases, although a possibility that it extends to a narrow gray hatched region in Fig.~\ref{fig:phasediagram}(b) has not been ruled out.  

v) Even-parity dimer (D$_-$) state ---
This has $\langle \hat{D}^x\rangle\langle \hat{D}^z\rangle<0$, while the vector spin chirality eventually vanishes, i.e., $\langle\hat{\kappa}^z\rangle=0$. 

The D$_\pm$ phases belong to the same symmetry group $G$ as that of the Hamiltonian, $G_{\mathscr{H}}$, which contains U(1) for the spin symmetry, the group $T$ of translations by integer multiples of two sites, the dihedral point group $D_{2h}=D_2\times C_1$ with $C_1=\{E,I\}$ and the spatial inversion $I$ about a bond center, and the anti-unitary group $\{E,\Theta\}$ with the identity $E$ and the time-reversal $\Theta$. The VCD$_\pm$ and VCD$_0$ also have a common symmetry group $G_{\mathrm{VCD}}$, which can be derived by replacing $D_{2h}$ with $C_{2v}$ where the inversion symmetry is lost while two mirror planes are preserved.
Clearly, the D$_+$---VCD$_+$ and D$_-$---VCD$_-$ transitions are symmetry-breaking transitions, which belong to the Ising criticality described with the $c=1/2$ CFT~\cite{UO14}. In particular, it breaks the $I$ symmetry while preserving the mirror symmetry including the $z$ axis, e.g., $IR_{2x}$ with the $\pi$ rotation $R_{2i}$ about the $i$ axis. In contrast, the VCD$_+$---VCD$_-$ transition is not if it occurs as a direct transition. 
We probe this VCD$_+$---VCD$_-$ transition only from the sign change of $\langle\hat{D}^x\rangle\langle\hat{D}^z\rangle$ but also from two string order parameters~\cite{denNijs89,Tasaki91} ${O}^z_n~(n=1,2)$ defined by 
\begin{equation}
{O}^z_n  \! = \! - \! \lim_{r\to\infty}\langle(\hat{S}^z_{n} \! + \! \hat{S}^z_{n+1}) e^{ i \pi \! \sum^{2r+n-1}_{k=n+2} \! \hat{S}^z_k } 
(\hat{S}^z_{2r+n} \! + \! \hat{S}^z_{2r+n+1})\rangle.
\label{string}
\end{equation}
Only $\langle O^z_s \rangle$ ($\langle O^z_w \rangle$) with a pair of sites 
$n$ and $n+1$ 
belonging to different dimer units (see Table \ref{table}) and thus forming a strong (weak) bond becomes long-range in the D$_{+(-)}$ and VCD$_{+(-)}$ phases, as shown in Fig.~\ref{fig:phasediagram}(b) and in the previous work~\cite{UO14}. 

\begin{table*}[bth]
  \caption{ (Color online) 
Ten $Z_2$ indices for the projective representation of $G_{\mathscr{H}}$ in D$_\pm$, VCD$_\pm$, and VCND~\cite{UO14} ground states, the degeneracy $n_s$/$n_w$ of the lowest entanglement spectrum $\zeta_0 = - \log w_0$ and the schematic picture of the ground state of $\hat{\mathscr{H}}_s$/$\hat{\mathscr{H}}_w$ when dividing the system at a stronger/weaker (left/right panel) bond. The emergence of $-1$ in $\beta$, $\gamma$ and/or $\omega$ points to a double topological degeneracy in the lowest entanglement spectrum. 
Orange, green and pink pairs indicate antisymmetric [$(|\! \uparrow\downarrow\rangle-|\downarrow\uparrow\rangle)/\sqrt{2}$], symmetric [$(|\! \uparrow\downarrow\rangle+|\downarrow\uparrow\rangle)/\sqrt{2}$] and mixed [$( e^{\i \theta/2} |\! \uparrow\downarrow\rangle+e^{-\i \theta/2}|\! \downarrow\uparrow\rangle)/\sqrt{2}$] units of dimers which show $\langle \hat{D}^{z}_j \rangle \langle \hat{D}^{x}_j \rangle > 0$ and $\langle \hat{D}^{z}_j \rangle \langle \hat{D}^{x}_j \rangle < 0$, respectively. These parity symmetries are broken in pink pairs due to presence of vector-chiral order. 
The twofold Kramers degeneracy arising from the edge is denoted by a pair of black up and down arrows. 
}
\label{table}
\begin{tabular}{c|c|c|c|c|c|c|c|c|c|c|c|cc|cc}
\hline \hline
 & & & \multicolumn2{c|}{$\alpha(h)$} & & & \multicolumn2{c|}{$\gamma(p,h)$} & \multicolumn2{c|}{$\gamma(\Theta,h)$} & & \multicolumn4{c}{Degeneracy $n_s$/$n_w$ of the ground state of $\hat{\mathscr{H}}_s$/$\hat{\mathscr{H}}_w$} \\
\cline{4-5}\cline{8-11}\cline{13-16}
Phase & $p$ & $\alpha(p)$ & $R_{2x}$ & $R_{2z}$ & $\beta(p)$ & $\beta(\Theta)$ & $R_{2x}$ & $R_{2z}$ & $R_{2x}$ & $R_{2z}$ & $\omega(R_{2x},R_{2z})$ & $n_{\mathrm{s}}$ &
\begin{minipage}{32mm}
\scalebox{0.38}{\includegraphics{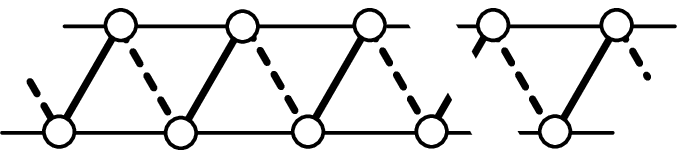}}
\end{minipage} 
& $n_{\mathrm{w}}$ &
\begin{minipage}{32mm}
\scalebox{0.38}{\includegraphics{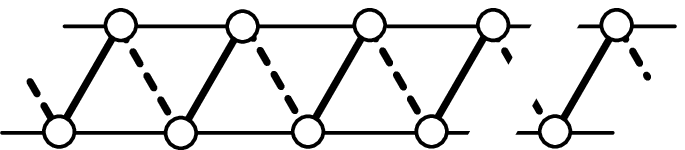}}
\end{minipage}   \\
\hline
D$_+$ & $I$ & $-1$ & $+1$ & $+1$ & $\pm1$ & $\pm1$ & $\pm1$ & $\pm1$ & $\pm1$ & $\pm1$ & $\pm1$ & 1 &
\begin{minipage}{32mm}
\scalebox{0.38}{\includegraphics{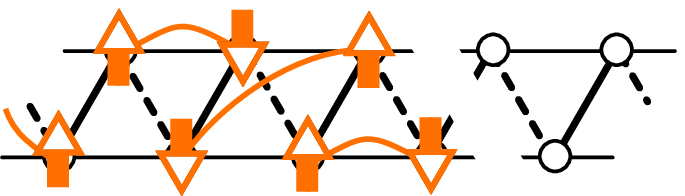}}
\end{minipage} 
& 2 &
\begin{minipage}{32mm}
\scalebox{0.38}{\includegraphics{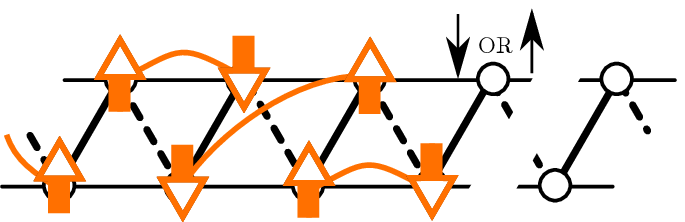}}
\end{minipage}   \\
\hline
D$_-$ & $I$ & $+1$ & $-1$ & $+1$ & $+1$ & $\mp1$ & $+1$ & $\mp1$ & $\mp1$ & $\mp1$ & $\mp1$ & 2 &
\begin{minipage}{32mm}
\scalebox{0.38}{\includegraphics{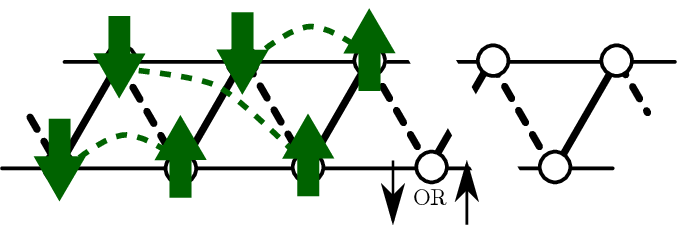}}
\end{minipage} 
& 1 &
\begin{minipage}{32mm}
\scalebox{0.38}{\includegraphics{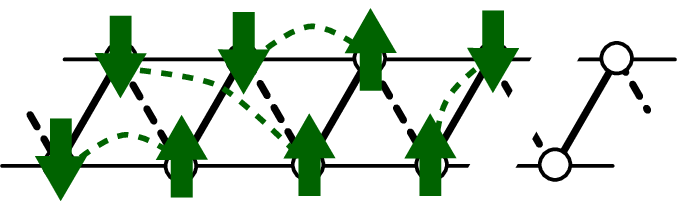}}
\end{minipage}   \\
\hline
VCD$_+$ & $IR_{2x}$ & $-1$ & $0$ & $+1$ & $+1$ & $\pm1$ & $0$ & $+1$ & $0$ & $\pm1$ & $0$ & 1 &
\begin{minipage}{32mm}
\scalebox{0.38}{\includegraphics{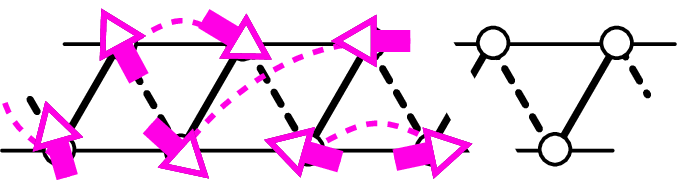}}
\end{minipage} 
& 2 &
\begin{minipage}{32mm}
\scalebox{0.38}{\includegraphics{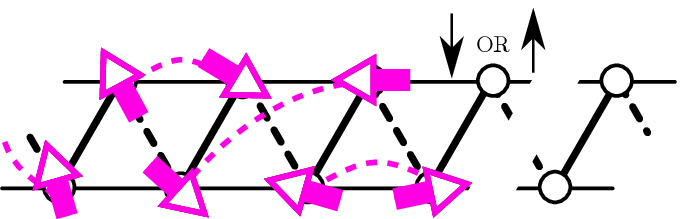}}
\end{minipage}   \\
\hline
VCD$_-$ & $IR_{2x}$ & $-1$ & $0$ & $+1$ & $+1$ & $\mp1$ & $0$ & $+1$ & $0$ & $\mp1$ & $0$ & 2 &
\begin{minipage}{32mm}
\scalebox{0.38}{\includegraphics{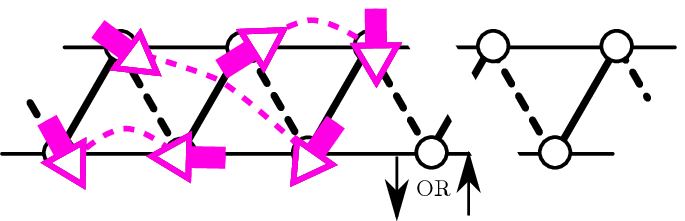}}
\end{minipage} 
& 1 &
\begin{minipage}{32mm}
\scalebox{0.38}{\includegraphics{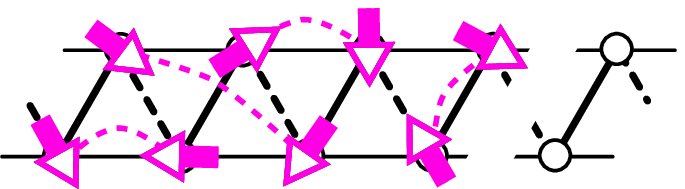}}
\end{minipage}   \\
\hline
VCND  & $IR_{2x}$ & $-1$ & $0$ & $+1$ & $+1$ & $0$ & $0$ & $+1$ & $0$ & $0$ & $0$ & 1 &
\begin{minipage}{32mm}
\scalebox{0.38}{\includegraphics{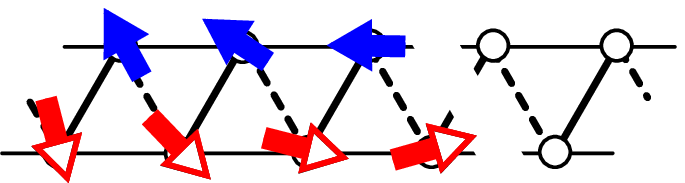}}
\end{minipage} 
& 1 &
\begin{minipage}{32mm}
\scalebox{0.38}{\includegraphics{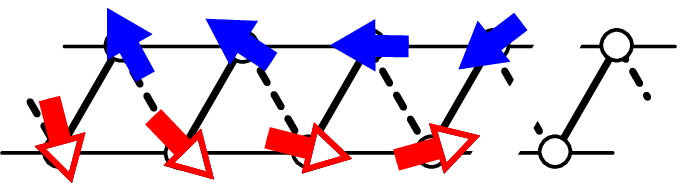}}
\end{minipage}   \\
\hline \hline
\end{tabular}
\end{table*}

This change in the string order parameters is consistent with a change in the degeneracy of the lowest entanglement spectrum. In two rightmost columns of Table~\ref{table}, we show the degeneracy $n_s$ ($n_w$) of the lowest bipartite entanglement spectrum, or in other words, that of the entanglement Hamiltonian~\cite{Li08} $\hat{\mathscr{H}}_s$ ($\hat{\mathscr{H}}_w$) obtained through iDMRG calculations under the condition that the whole spin chain is divided at a strong (weak) bond:  $n_s=1$ and $n_w=2$ for the D$_+$ and VCD$_+$ phases, while $n_s=2$ and $n_w=1$ for the D$_-$ and VCD$_-$ phases. This topological change occurring only at the VCD$_+$---VCD$_-$ transition indicates that the $D_\pm$ phases are not adiabatically connected and neither are the VCD$_\pm$ phases, as long as the symmetry of these phases is respected. 

Nature of these gapped phases can be captured by classifying them as SPT phases, according to the 1D representations and the factor systems for the projective representation of the symmetry group $G$ of each ground state. Let us consider the set of $Z_2$ indices, $\alpha's$, $\beta's$, $\gamma's$, and $\omega's$ listed in Table \ref{table}~\cite{Chen11,Pollmann10,Pollmann12} for the symmetry group $G_\mathscr{H}$ of the Hamiltonian, so that the symmetry group $G$ of all the ground states of our interest can be given by a subgroup of $G_\mathscr{H}$. These indices are determined from
\begin{eqnarray}
\sum_{jj'} (T_I(h_p))_{ii',jj'} (D_{p})_{j,j'} & = & \alpha(p) (D_{p})_{i,i'}, \label{eq:ti} \\
\sum_{jj'} (T_\Theta(h_{\Theta}))_{ii',jj'} (D_{\Theta})_{j,j'} & = & \alpha(\Theta) (D_{\Theta})_{i,i'}, \label{eq:tt} \\
\sum_{jj'} (T(h))_{ii',jj'} (\mathcal{U}_{h})_{j,j'} & = & \alpha(h) (\mathcal{U}_{h})_{i,i'}, \label{eq:uh} 
\end{eqnarray}
\vspace{-4mm}
\begin{eqnarray}
\beta(p) & = & \Tr[D_{p} ( D^{-1}_{p} )^{\rm t}]/m,~
\beta(\Theta) = \Tr[D_{\Theta}D^*_{\Theta}]/m, 
\end{eqnarray}
\vspace{-6.5mm}
\begin{eqnarray}
\gamma(p,h) & = & \Tr[\mathcal{U}_{h}D_{p} \mathcal{U}^{\rm t}_{h} D^{-1}_{p}]/m,~\\
\gamma(\Theta,h) & = & \Tr[\mathcal{U}_{h}D_{\Theta}(\mathcal{U}^{*}_{h})^{-1}D^{-1}_{\Theta}]/m,~\\
\omega(h',h) & = & \Tr[\mathcal{U}_{h}\mathcal{U}_{h'}\mathcal{U}^{-1}_{h}\mathcal{U}^{-1}_{h'}]/m,~\label{eq:gamma}
\end{eqnarray}
where $h$ is taken from a minimal set of generators of the local unitary subgroup $H_{LU}$ of the whole symmetry group $G$, $p = Ih_p$ is a direct product of the inversion $I$ and $h_p=E$ or $R_{2x} \in H_{LU}$, $\Theta= h_\Theta K$ is a direct product of the complex conjugate operator $K$ and $h_\Theta=R_{2y}\in H_{LU}$. 
We have also introduced transfer matrices for a unit cell including two spins
\begin{eqnarray}
&&\hspace{-10mm} (T_{I}(h))_{ii',jj'} \hspace{-1mm} = \hspace{-4mm} 
\sum_{s_1 s_2 s_1' s_2'} \hspace{-3mm} (A^{*(s_1 s_2)})_{i j} (U_h)_{s_1 s_2 s_1' s_2'} (A^{{\rm t}(s_2' s_1')})_{i' j'} , \\
&&\hspace{-10mm} (T_{\Theta}(h))_{ii',jj'} \hspace{-1mm} = \hspace{-4mm}
\sum_{s_1 s_2 s_1' s_2'} \hspace{-3mm} (A^{*(s_1 s_2)})_{i j} (U_h)_{s_1 s_2 s_1' s_2'} (A^{*(s_1' s_2')})_{i' j'} , \\
&&\hspace{-10mm} (T(h))_{ii',jj'} \hspace{-1mm} = \hspace{-4mm}
\sum_{s_1 s_2 s_1' s_2'} \hspace{-3mm} (A^{*(s_1 s_2)})_{i j} (U_h)_{s_1 s_2 s_1' s_2'} (A^{(s_1' s_2')})_{i' j'} , \label{eq:th}
\end{eqnarray}
where the $m \times m$ matrix $A^{(s_1 s_2)}$ represents in the Schmidt bases the state within the translation unit having the two-spin degrees of freedom, $(s_1 s_2)$, in the translationally invariant matrix product state (MPS)~\cite{Ostlund95, Rommer97, Schollwock11} $\ket{\Psi}_i= \sum_{s_1s_2 j} (A^{(s_1s_2)})_{ij} \ket{s_1 s_2} \otimes \ket{\Psi}_j$ of the entanglement Hamiltonian satisfying the orthonormal condition $_i\braket{\Psi}{\Psi}_j = \delta_{ij}$~\cite{mps4to2}. 
Right eigenvectors of transfer matrices in Eqs. (\ref{eq:ti}), (\ref{eq:tt}), and (\ref{eq:uh}) are the representation matrices of  $I$, $\Theta$, and $h$, respectively, in the Schmidt bases. (See Supplementary materials.)
The arbitrary phases of $\mathcal{U}_{R_{2x}}$ and $\mathcal{U}_{R_{2z}}$ are fixed by $\mathcal{U}_{R_{2x}}^2 = \mathcal{U}_{R_{2z}}^2 = \1$. Note that for $\Theta$-invariant states, i.e, $|\alpha(\Theta)|=1$, $\alpha(\Theta)$ just takes arbitrary U(1) phase depending on that of
$A^{(s1s2)}$ and thus is not important. The results are summarized in Table~\ref{table}. Because of the unbroken U(1)$_z$ symmetry, the 1D representation $\alpha(R_{2z})=1$ leading to $R_{2z}$-even states is rather obvious in all the phases shown in Table~\ref{table}, and thus is not particularly mentioned below.

From two 1D representations $\alpha(I)$ and $\alpha(R_{2x})$, the D$_+$ ground state of the whole spin chain is $I$-odd and $R_{2x}$-even. All the other $Z_2$ indices take the same value; $\beta(I)=\beta(\Theta)=\gamma(I,h)=\gamma(\Theta,h)=\omega(R_{2x},R_{2z})=+(-)1$ with $h=R_{2x},R_{2z}$ if the spin chain is cut at a strong (weak) bond. This is consistent with the nondegeneracy $n_s=1$ and the twofold degeneracy $n_w=2$ in the entanglement spectrum, and indicates that this SPT phase is protected by $I$, $\Theta$, and $D_2$ symmetries~\cite{Pollmann10,Pollmann12}. This phase has the same $Z_2$ indices as the Affleck-Kennedy-Lieb-Tasaki (AKLT) state~\cite{Affleck87, mps_AKLT}. 

The D$_-$ phase is $I$-even ($\alpha(I)=+1$) and $R_{2x}$-odd ($\alpha(R_{2x})=-1$). Whichever bond the spin chain is cut at, $\beta(I)=\gamma(I,R_{2x})=+1$, indicating that the $I$ symmetry no longer protects the topological degeneracy. All the other indices take the same value; $\beta(\Theta)=\gamma(I,R_{2z})=\gamma(\Theta,h)=\omega(R_{2x},R_{2z})=-(+)1$ if the spin chain is cut at a strong (weak) bond. This is consistent with $n_s=2$ and $n_w=1$, and indicates that this SPT phase is protected by $\Theta$ and $D_2$ symmetries. 
This phase has the same $Z_2$ indices as a direct product of the even-parity dimer state, $(\ket{\uparrow\downarrow}+\ket{\downarrow\uparrow})/\sqrt{2}$~\cite{mps_MG}.

Let us proceed to the VCD$_\pm$ phases. These states respect the $IR_{2x}$ symmetry and are $IR_{2x}$-odd, while they break the $I$ and $R_{2x}$ symmetries, as seen from $\alpha(IR_{2x})=-1$ and $\alpha(R_{2x})=0$. (Note that the D$_\pm$ states are also $IR_{2x}$-odd as $\alpha(IR_{2x})=\alpha(I)\alpha(R_{2x})=-1$.) The VCD$_\pm$ states have the same topological degeneracy in the entanglement spectrum as D$_\pm$, respectively, but they are no longer protected by the $I$ and $D_2$ symmetry, and not even by the $IR_{2x}$ symmetry since $\beta(IR_{2x}) = \gamma(IR_{2x},R_{2z}) = +1$ always holds. The sign of $\beta(\Theta)=\gamma(\Theta,R_{2z})$ depends on the way of dividing the spin chain and are opposite between the VCD$_+$ an VCD$_-$ phases, and the minus sign appears when the entanglement spectrum is twofold degenerate. Hence, VCD$_\pm$ phases are classified into distinct SPT phases, whose distinction is protected by the $\Theta$ symmetry. Indeed, once, the Neel LRO is realized in addition to the VCD orders, the $\Theta$ symmetry is broken~\cite{UO14} and the topological degeneracy is lost completely (Table \ref{table}).

\begin{figure}
\includegraphics[width=8.6cm]{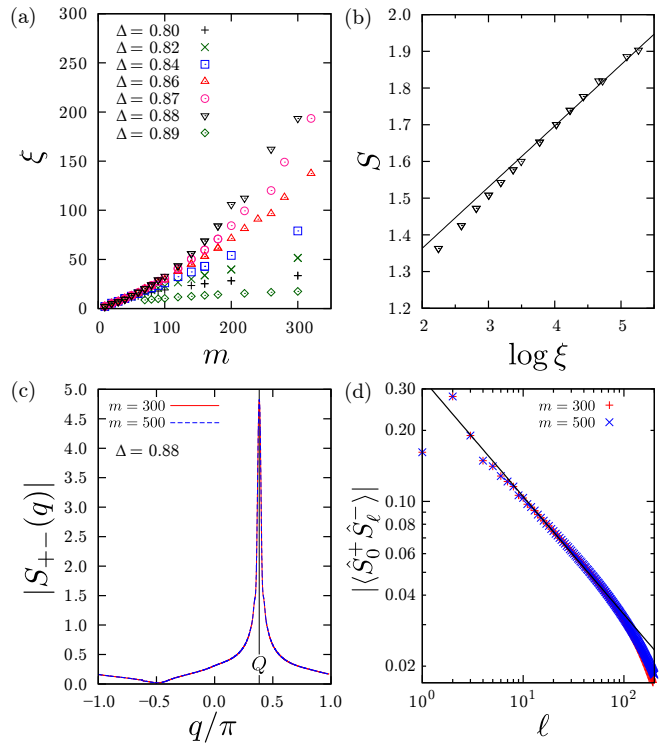}
  \caption{ 
(Color online) (a) Correlation length $\xi$ as a function of $m$ in the vicinity of the VCD$_+$---VCD$_{-}$ phase boundary. (b) Scaling of the entanglement entropy $S$ as a function of correlation length $\xi$ at $\Delta = 0.88$. The solid line represents the $c=1$ line. (c) The Fourier transform $S_{+-}(q)$ of $\langle \hat{S}^{+}_{0} \hat{S}^{-}_{\ell}\rangle$ ($1 \leq \ell \leq 128$). It exhibits a peak at $q = Q$ with $Q/\pi = 0.383$ denoted by the solid line. (d) Logarithmic plot of $|\langle \hat{S}^{+}_{0} \hat{S}^{-}_{\ell}\rangle|.$ The solid curve shows the scaling function given in the text with $A = 0.332(4)$ and $K_+=1.00(1)$, where the number in a parenthesis means the standard error coming from the least-square fitting in the range $5 \leq \ell \leq 100$. The downward deviation for $\ell>\xi\sim100-200$ is due to the effect of the truncation.
}
\label{fig:scaling}
\end{figure}

Finally, we clarify the nature of this VCD$_{+}$---VCD$_{-}$ SPT phase transition. Figure~\ref{fig:scaling}~(a) shows the dependence of the correlation length $\xi =-1/\log (|w_1/w_0|)$ on the dimension $m$ of the Schmidt bases in the vicinity of VCD$_+$---VCD$_-$ transition, where $w_n$ is the $(n+1)$th-largest (in terms of absolute value) eigenvalue of the transfer matrix $T(E)$. This indicates the strongest enhancement of $\xi$ at $\Delta=0.88$, indicating a proximity to the criticality in reasonable agreement with the sign change of $\langle \hat{D}^z \rangle$ at $\Delta=0.879(1)$. The scaling behavior of the entanglement entropy versus $\xi$ in the form of $S=\frac{c}{6} \log \xi + {\rm const.}$ shown in Fig.~\ref{fig:scaling} (b) is consistent, within the numerical accuracy, with the $c=1$ CFT~\cite{CFT96}. We also estimate the Tomonaga-Luttinger (TL) parameter $K_{+}$ for the gapless VCD$_{0}$ state~\cite{FSOF12, UO14} to be unity (1.00(1)), the same value as for the TL liquid in the SU(2) NN antiferromagnetic spin-$\frac{1}{2}$ chain,  by fitting a spatial decay of the transverse equal-time spin correlation with the leading term as 
$\langle \hat{S}^x_{0} \hat{S}^x_{\ell} \rangle \simeq A e^{\i Q\ell} |\ell|^{-1/(2K_{+})}$, 
as shown in Fig. \ref{fig:scaling}~(c) and (d).
If we applying the heuristic bosonization analysis~\cite{Nersesyan98} to our model~\cite{UO14}, this value $K_{+} = 1$ is indeed required for having a direct continuous transition between the VCD$_{\pm}$ phases~\cite{UO14}. This supports the scenario that two Berezinskii-Kosterlitz-Thouless (BKT) transitions at $K_+=\frac{1}{2}$ from the gapless VC to gapped VCD$_\pm$ phases in the case of $\delta=0$ shift and meet each other at the $K_+=1$ line in the case of $\delta\ne0$ (see Fig.1~(a)): the change of the critical $K_+$ value is caused by an appearance of the more relevant perturbation of the bond alternation~\cite{UO14}. This contrasts to the case of the transition between the large-$D$ and Haldane phases, which has a simple Gaussian criticality with a weak universality~\cite{Yamanaka93,Hida93}. Analytically describing the possible coincidence of two BKT transitions at the $K_+=1$ Gaussian criticality is left open.

Since the model parameters are at least close to those for the spin-gapped spin-$1/2$ chain compound Rb$_2$Cu$_{2}$Mo$_{4}$O$_{12}$~\cite{UO14,Hase04,Yasui13}, it would be intriguing to experimentally find these SPT phases and the SPT transition by probing a gap closing under physical and/or chemical pressure.

%
%%%%%%%%%%%%%%%%%%%%%%%%%%%%%%%%%%%%%%%%%%%
%\acknowledgement 

The authors thank F. Pollmann, S. Furukawa, and M. Oshikawa  for stimulating discussions. The work was partially supported by Grants-in-Aid for Scientific Research under Grant No. 24740253 (S.O.) and 25800221 (H.U.) from Japan Society for the Promotion of Science, by the RIKEN iTHES Project, and by the National Science Foundation under Grant No. PHYS-1066293 and the hospitality of the Aspen Center for Physics (S.O.).

\newcommand{\PRL}[3]{Phys.\ Rev.\ Lett.\ {\bf #1}, #2 (#3)}
\newcommand{\PRLp}[3]{Phys.\ Rev.\ Lett.\ {\bf #1}, #2 (#3)}
\newcommand{\PRA}[3]{Phys.\ Rev.\ A {\bf #1}, #2 (#3)}
\newcommand{\PRAp}[3]{Phys.\ Rev.\ A {\bf #1}, #2 (#3)}
\newcommand{\PRB}[3]{Phys.\ Rev.\ B {\bf #1}, #2 (#3)}
\newcommand{\PRBp}[3]{Phys.\ Rev.\ B {\bf #1}, #2 (#3)}
\newcommand{\PRBR}[3]{Phys.\ Rev.\ B {\bf #1}, #2 (R) (#3)}
\newcommand{\PRBRp}[3]{Phys.\ Rev.\ B {\bf #1}, R#2 (#3)}
\newcommand{\arXiv}[1]{arXiv:#1}
\newcommand{\condmat}[1]{cond-mat/#1}
\newcommand{\JPSJ}[3]{J. Phys.\ Soc.\ Jpn.\ {\bf #1}, #2 (#3)}
\newcommand{\PTPS}[3]{Prog.\ Theor.\ Phys.\ Suppl.\ {\bf #1}, #2 (#3)}


\begin{thebibliography}{9}

\bibitem{Chen10}
%Local unitary transformation, long-range quantum entanglement, wave function renormalization, and topological order
  X. Chen, Z.-C. Gu, and X.-G. Wen, Phys. Rev. B \textbf{82}, 155138 (2010).

\bibitem{Wen02}
%Quantum orders and symmetric spin liquids. 
  X.-G. Wen, Phys. Rev. B \textbf{65}, 165113 (2002).

\bibitem{Hasan10}
  M. Z. Hasan and C. L. Kane, Rev. Mod. Phys. \textbf{82}, 3045 (2010).

%[Haldane chain]
\bibitem{Haldane83}
 F.D.M. Haldane, Phys. Lett. 93A, 464 (1983); \PRLp{50}{1153}{1983}. 
\bibitem{Affleck87}
 I. Affleck, T. Kennedy, E. H. Lieb, and H. Tasaki, \PRLp{59}{799}{1987}. 

%Symmetry operator
\bibitem{Pollmann10}
F. Pollmann, A.M. Turner, E. Berg, and M. Oshikawa, \PRB{81}{064439}{2010}.
\bibitem{Pollmann12}
F. Pollmann and A.M. Turner, \PRB{86}{125441}{2012}.

\bibitem{Liu11}
%Symmetry-protected topological orders of one-dimensional spin systems with D2 + T symmetry
  Z.-X. Liu, X. Chen, and X.-G. Wen, Phys. Rev. B \textbf{84}, 195145 (2011).
\bibitem{Chen11} 
%Complete classification of one-dimensional gapped quantum phases in interacting spin systems
  X. Chen, Z.-C. Gu, and X.-G. Wen, Phys. Rev. B \textbf{84}, 235128 (2011).
\bibitem{Gu13}
%Symmetry protected topological orders and the group cohomology of their symmetry
  X. Chen, Z.-C. Gu, Z.-X. Liu, and X.-G. Wen, Science \textbf{338}, 1604 (2012); Phys. Rev. B \textbf{87}, 155114 (2013).
 
\bibitem{Schuch11}
N. Schuch, D. P\'erez-Garc{\ia}a, and I. Cirac, Phys. Rev. B \textbf{84}, 165139 (2011).
 
\bibitem{Liu12}
%Symmetry-protected topological phases in spin ladders with two-body interactions
  Z.-X. Liu, Z.-B. Yang, Y.-J. Han, W. Yi, and X.-G. Wen, Phys. Rev. B \textbf{86}, 195122 (2012).

\bibitem{Chubukov91}
 A.V. Chubukov, \PRB{44}{4693}{1991}. 

\bibitem{Lecheminant05}
 P. Lecheminant, in {\it Frustrated spin systems}, 
 edited by H. T. Diep (World-Scientific, Singapore, 2005), Review chapter;
 \arXiv{cond-mat/0306520}.  

%gapless chiral phase
\bibitem{Nersesyan98}
  A.A. Nersesyan, A.O. Gogolin, and F.H.L. E{\ss}ler, \PRL{81}{910}{1998}.

\bibitem{Hikihara01}%S=1/2,3/2,2
  T. Hikihara, M. Kaburagi, and H. Kawamura, \PRB{63}{174430}{2001}.

\bibitem{FSO10}
  S. Furukawa, M. Sato, and S. Onoda, \PRL{105}{257205}{2010}.
%phase diagram
\bibitem{FSOF12}
  S. Furukawa, M. Sato, S. Onoda, and A. Furusaki, Phys. Rev. B \textbf{86}, 094417 (2012).

%[spin-1/2 multiferroics]
%(LiCu2O2)
\bibitem{Masuda05}
 T. Masuda, A. Zheludev, B. Roessli, A. Bush, M. Markina, and A. Vasiliev,  
 \PRB{72}{014405}{2005}.
\bibitem{Park07}
   S. Park, Y. J. Choi, C. L. Zhang, and S-W. Cheong,
   \PRL{98}{057601}{2007}.

%(LiCuVO4)
\bibitem{Enderle05}
  M.~Enderle \textit{et al.},
%, C.~Mukherjee, B.~F{\aa}k, R.K.~Kremer, J.-M.~Broto, H.~Rosner, S.-L.~Drechsler, J.~Richter, J.~Malek, A.~Prokofiev, W.~Assmus, P. Pujol, J.-L.~Raggazzoni, H.~Rakoto, M.~Rheinst\"adter, and H.M.~R{\o}nnow, 
  Europhys.\ Lett.\ \textbf{70}, 237 (2005).
\bibitem{Naito07}
  Y. Naito, K. Sato, Y. Yasui, Y. Kobayashi, Y. Kobayashi, and M. Sato,
  \JPSJ{76}{023708}{2007}.

%(PbCuSO4(OH)2)
\bibitem{Yasui11}
Y. Yasui, M. Sato, and I. Terasaki, 
\JPSJ{80}{033707}{2011}.
\bibitem{Wolter012}
A. U. B. Wolter, F. Lipps, M. Sch\"apers, S.-L. Drechsler, S. Nishimoto, 
R. Vogel, V. Kataev, B. B\"uchner, H. Rosner, M. Schmitt, M. Uhlarz, 
Y. Skourski, J. Wosnitza, S. S\"ullow, and K. C. Rule, 
\PRB{85}{014407}{2012}.

%[J1-J2 chains showing no magnetic LRO, phase diagram]
\bibitem{UO14}
  H. Ueda and S. Onoda, Phys. Rev. B \textbf{89}, 024407 (2014).

%[J1-J2 chains showing no magnetic LRO]
\bibitem{Hase04}
 M. Hase, H. Kuroe, K. Ozawa, O. Suzuki, H. Kitazawa, G. Kido, and T. Sekine, 
 \PRB{70}{104426}{2004}.
\bibitem{Yasui13}
  Y. Yasui, Y. Yanagisawa, R. Okazaki, I. Terasaki, Y. Yamaguchi, and T. Kimura, J. Appl. Phys. \textbf{113}, 17D910 (2013)

%iDMRG
\bibitem{iDMRG}
I. P. McCulloch, \arXiv{0804.2509}

\bibitem{CFT96}
P.~Di~Francesco, P~Mathieu, and D.~S\'{e}n\'{e}chal, \textit{Conformal Field Theory} (Springer, New York, 1996). 

\bibitem{Itoi01}
  C. Itoi and S. Qin, \PRB{63}{224423}{2001}.

%[String order parameter]
\bibitem{denNijs89}
 M. den Nijs and K. Rommelse, \PRBp{40}{4709}{1989}. 
\bibitem{Tasaki91}
 H. Tasaki, \PRLp{66}{798}{1991}. 
 
 %Entanglement Hamiltonian
\bibitem{Li08} Hui~Li and F.~D.~M.~Haldane, \PRL{101}{010504}{2008}.
 
 %MPS
\bibitem{Ostlund95}
S. \"Ostlund and S. Rommer, \PRL{75}{3537}{1995}.
\bibitem{Rommer97}
S. Rommer and S. \"Ostlund, \PRB{55}{2164}{1997}.
\bibitem{Schollwock11}
U. Schollw\"ock, Ann. Phys. (NY) {\bf 326}, 96 (2011).
 
\bibitem{mps4to2} Actually, we first adopted an infinite MPS $\ket{\Psi}$ invariant under the four-site translation in our iDMRG calculation. Then, we checked the state is invariant under two-site translations, namely, $\ket{\Psi} = \hat{T} \ket{\Psi}$.
 
\bibitem{mps_AKLT} The AKLT state on the $S=1/2$ chain is described by a product state of the following translation-unit matrix $A^{(s_1 s_2)}_{\rm w}$ or $A^{(s_1 s_2)}_{\rm s}$ in the Schmidt bases obtained by dividing the whole system at a weak or strong $J_1$ bond; $A^{(\uparrow \uparrow)}_{\rm w} = \sqrt{\frac{2}{3}}\sigma^{+}$, $A^{(\downarrow \downarrow)}_{\rm w} = -\sqrt{\frac{2}{3}}\sigma^{-}$ and $A^{(\uparrow \downarrow)}_{\rm w} = A^{(\downarrow \uparrow)}_{\rm w} = -\sqrt{\frac{1}{6}}\sigma^z$. $A^{(s_1s_2)}_{\rm s}$ is readily obtained by applying the singular value decomposition to $A^{(s_1s_2)}_{\rm w}$; using singular vectors and values in $(A^{(s_1 s_2)}_{\rm w})_{\alpha_1 \alpha_2} = \sum_{\beta = 1}^{4} X_{\alpha_1 s_1,\beta} W_{\beta,\beta} Y^{\dagger}_{\beta,s_2 \alpha_2}$, we can take $(A^{(s_1s_2)}_{\rm s})_{\beta_1\beta_2}=\sum_{\alpha} W_{\beta_1,\beta_1} Y^{\dagger}_{\beta_1,s_1\alpha} X_{\alpha s_2,\beta_2}$.

\bibitem{mps_MG}
$A^{(s_1s_2)}_{\rm w}$ of the direct product state of $(\ket{\!\! \uparrow\downarrow}+\ket{\!\! \downarrow\uparrow})/\sqrt{2}$ can be written as  $A^{(\uparrow\uparrow)}_{w} = A^{(\downarrow\downarrow)}_{w} = \begin{pmatrix}
0 \\
\end{pmatrix}$, 
$A^{(\uparrow\downarrow)}_{w} = A^{(\downarrow\uparrow)}_{w} = \begin{pmatrix}
\frac{1}{\sqrt{2}} \\
\end{pmatrix}$.
$A^{(s_1s_2)}_{\rm s}$ of this state can be obtained in the same way as explained in \cite{mps_AKLT}. These $A^{(s_1s_2)}_{w/s}$ lead to the same $Z_2$ indices as in the D$_{-}$ state. 
$A^{(s_1s_2)}_{w/s}$ of the direct product state of $(\ket{\!\! \uparrow\downarrow}-\ket{\!\! \downarrow\uparrow})/\sqrt{2}$, namely, the Majumdar-Ghosh state~\cite{Majumdar69}, can be obtained by rotating every other spins about the $z$-axis by $\pi$. This resulting $Z_2$ indices are the same as in the D$_+$ state, except with the interchange of ``w" and ``s".

\bibitem{Majumdar69}
C. K. Majumdar and D. K. Ghosh, J. Math. Phys. \textbf{10}, 1399 (1969).

\bibitem{Yamanaka93}
  M. Yamanaka, Y. Hatsugai, and M. Kohmoto, Phys. Rev. B \textbf{48}, 9555 (1993).

\bibitem{Hida93}
  K. Hida, J. Phys. Soc. Jpn. \textbf{62}, 1466 (1993).

\end{thebibliography}
\end{document}

\onecolumngrid

\section{Supplementary materials for ``Symmetry-protected topological phases and transition in a frustrated spin-$\frac{1}{2}$ XXZ chain"}

We provide details of a correspondence between the right eigenvector of the transfer matrix given by Eq. (5) in the main text and the representation matrix in the Schmidt bases. 
Schmidt bases of a finitely correlated ground state for a uniform one-dimensional system in the thermodynamic limit can be represented by an infinite matrix product state with sufficiently large dimension $m$: 
\begin{equation}
\ket{\Psi}_i = \sum_{ \{s_k \} } \Big[ \big( \prod_{k = 1}^{+\infty} A^{(s_k)} \big) v \Big]_i \ket{\{ s_k \} },
\end{equation}
where $A^{(s)}$ and $v$ are an $m$-dimensional matrix and vector. 
The gauge of $A^{(s)}$ can be chosen to be $\sum_{s} A^{(s)} A^{\dagger (s)} = \1$.
The vector $v$ is determined to suit the orthonormal condition, $_i\braket{\Psi}{\Psi}_j=\delta_{ij}$. 
The bases have a translation symmetry given by $\ket{\Psi}_i = \sum_{s} (A^{(s)})_{ij} \ket{s} \otimes \ket{\Psi}_j$. 

Let's consider a representation matrix of a local unitary operation, $\hat{U} = \prod_{k=1}^{+\infty} \sum_{s_ks'_k} (U_h)_{s_ks'_k} \ket{s_k}\bra{s'_k}$, in the Schmidt bases, namely $(\mathcal{U}_h)_{ii'} = _i \!\! \bra{\Psi} \hat{U} \ket{\Psi}_{i'}$: 
\begin{eqnarray}
_i \bra{\Psi} \hat{U} \ket{\Psi}_{i'} & = & \sum_{jj'} \Big[ \prod_{k=1}^{+\infty} \big( \sum_{s_k s'_k} (U_{h})_{s_k s'_k} A^{*(s_k)} \otimes A^{(s_k')} \big) \Big]_{ii',jj'} (v^{*})_j (v)_{j'}, \nonumber \\
& = & \sum_{jj'} \big[ \prod_{k=1}^{+\infty} T(h) \big]_{ii',jj'} (v^{*})_j (v)_{j'}, 
\end{eqnarray}
where the definition of the transfer matrix $T(h)$ is given by Eq. (12) in the main text.
If the ground state is invariant under the unitary operation and is not a cat state, the norm of dominant eigenvalue $\alpha(h)$ of the transfer matrix $T(h)$ becomes unity and unique. In this case, $\prod_{k=1}^{+\infty} T(h)$ can be decomposed as $u_h (\prod_{k=1}^{+\infty} \alpha(h)) v^{\dagger}_h$, where $u_h$ ($v_h$) is the right (left) eigenvector of $T(h)$ corresponding to $\alpha(h)$. Using this relation, we obtain
\begin{eqnarray}
_i \bra{\Psi} \hat{U} \ket{\Psi}_{i'} & = & (u_h)_{ii'} \big( \prod_{k=1}^{+\infty} \alpha(h) \big) \Big[  \sum_{jj'} (v_h^*)_{jj'} (v^{*})_j (v)_{j'} \Big] = (u_h)_{ii'} \times {\rm const.}, 
\end{eqnarray}
where the constant is $\big( \prod_{k=1}^{+\infty} \alpha(h) \big) \Big[  \sum_{jj'} (v_h^*)_{jj'} (v^{*})_j (v)_{j'} \Big]$. This constant can be removed by redefining of $u_h$ and $v_h$, because there is an arbitrary property in the biorthogonal condition of $v_h^{\dagger} u_h = 1$.
Thus, we can obtain the representation matrix by reshaping the right eigenvector, as $(\mathcal{U}_h)_{i,i'} = (u_h)_{ii'}$.

% \newcommand{\PRL}[3]{Phys.\ Rev.\ Lett.\ {\bf #1}, #2 (#3)}
% \newcommand{\PRLp}[3]{Phys.\ Rev.\ Lett.\ {\bf #1}, #2 (#3)}
% \newcommand{\PRA}[3]{Phys.\ Rev.\ A {\bf #1}, #2 (#3)}
% \newcommand{\PRAp}[3]{Phys.\ Rev.\ A {\bf #1}, #2 (#3)}
% \newcommand{\PRB}[3]{Phys.\ Rev.\ B {\bf #1}, #2 (#3)}
% \newcommand{\PRBp}[3]{Phys.\ Rev.\ B {\bf #1}, #2 (#3)}
% \newcommand{\PRBR}[3]{Phys.\ Rev.\ B {\bf #1}, #2 (R) (#3)}
% \newcommand{\PRBRp}[3]{Phys.\ Rev.\ B {\bf #1}, R#2 (#3)}
% \newcommand{\arXiv}[1]{arXiv:#1}
% \newcommand{\condmat}[1]{cond-mat/#1}
% \newcommand{\JPSJ}[3]{J. Phys.\ Soc.\ Jpn.\ {\bf #1}, #2 (#3)}
% \newcommand{\PTPS}[3]{Prog.\ Theor.\ Phys.\ Suppl.\ {\bf #1}, #2 (#3)}
% \begin{thebibliography}{9}
% \bibitem{Nersesyan98}
%   A.A. Nersesyan, A.O. Gogolin, and F.H.L. E{\ss}ler, \PRL{81}{910}{1998}.
% \bibitem{Hikihara08}
%  T. Hikihara, L. Kecke, T. Momoi, and A. Furusaki, \PRB{78}{144404}{2008}. 
% \bibitem{FSOF12}
%   S. Furukawa, M. Sato, S. Onoda, and A. Furusaki, Phys. Rev. B \textbf{86}, 094417 (2012).
% \bibitem{UO14}
%   H. Ueda and S. Onoda, Phys. Rev. B \textbf{89}, 024407 (2014).
% \bibitem{White:PRL69_White:PRB48} 
%  S.~R.~White, Phys. Rev. Lett. {\bf 69}, 2863 (1992); Phys. Rev. B {\bf 48},  10345 (1993).
% \bibitem{iDMRG}
% I. P. McCulloch, \arXiv{0804.2509}
% \end{thebibliography}